\documentclass{article}
\usepackage[margin=25mm]{geometry}
\usepackage[utf8]{inputenc}
\usepackage[usenames]{color}   
\usepackage[T1]{fontenc}    
\usepackage{hyperref}       
\usepackage{url}            
\usepackage{booktabs}       
\usepackage{amsfonts}       
\usepackage{amsmath}
\usepackage{nicefrac}       
\usepackage[nopatch=eqnum]{microtype}      
\usepackage{lipsum}
\usepackage{graphicx}
\usepackage{rotating}  
\usepackage{authblk}
\usepackage{cite} 
\providecommand{\keywords}[1]
{
  \small	
  \textbf{\textit{Keywords---}} #1
}

\hypersetup{
colorlinks=true,       
linkcolor=red,          
citecolor=blue,        
filecolor=magenta,      
urlcolor=red   
}
\DeclareUnicodeCharacter{2009}{\,} 

\newcommand{\NA}{\mathrm{NA}}
\newcommand{\R}{\mathrm{R}}
\newcommand{\FOV}{\mathrm{FOV}}
\newcommand{\eff}{\text{eff}}
\renewcommand{\k}{\mathrm{k}}
\renewcommand{\r}{\mathrm{r}}
\newcommand\norm[1]{\left\lVert#1\right\rVert}

\title{Illumination strategies for space-bandwidth-time product improvement in Fourier ptychography}

\author[1]{Haibo Xu}
\author[1]{Cheng Li}
\author[1]{Mingzhe Wei}
\author[1]{Ziwen Zhou}

\author[1,2]{Longqian Huang\thanks{ALL authors contributed equally}\thanks{Corresponding Author: longqianh@zju.edu.cn}}
\affil[1]{\normalsize{Institute of Laser Biomedicine, Zhejiang University, Hangzhou 310027, China}}

\affil[2]{\normalsize{School of Brain Science and Brain Medicine, Zhejiang University, Hangzhou 310058, China}}

\date{\vspace{-5ex}}
\begin{document}

\maketitle

\begin{abstract}
    Fourier ptychography (FP) is a promising technique for high-throughput imaging. Reconstruction algorithms and illumination paradigm are two key aspects of FP. In this review, we mainly focus on illumination strategies in FP. We derive the space-bandwidth-time product (SBP-T) for the characterization of FP performance. Based on the analysis of SBP-T, we categorize the illumination strategy in FP effectively and discuss each category in detail.

\end{abstract}
\hspace{10pt}


\section{Introduction}
Resolution and imaging field-of-view (FOV) are two factors that are mostly considered in imaging systems, especially in microscopy. Researchers have to choose microscopes with appropriate resolution and FOV for specific applications since there is a trade-off between them: high-resolution systems tend to have small FOV, and vice versa. Fourier ptychography (FP) is a synthetic approach that presumes high-throughput and high-resolution imaging by using more information through multiple captures. It captures multiple low-resolution, large FOV images and computationally combines them in the Fourier domain into a high-resolution, large FOV result \cite{konda2020fourier}. 

The first demonstration of FP was made in 2013 by Zheng \textit{et al.} \cite{zheng2013wide}. With a programmable LED matrix, they proposed an iterative phase-retrieval-like ptychography algorithm in the Fourier domain and achieved a half-pitch resolution of 0.78 mm, a FOV of $\sim 120 \ \text{mm}^2$ and a resolution-invariant depth of focus of 0.3 mm. Since then, FP has been developed in various directions, including higher resolution \cite{ou2015high,pacheco2016reflective,sun2017resolution,pan2018subwavelength,zhang2019near}, faster imaging speed \cite{dong2014sparsely,kuang2015digital,kappeler2017ptychnet,zhou2017fourier,sun2018single,he2018single,nguyen2018deep,cheng2019illumination,zhang2019fourier,sun2018high,xiao2021high,aidukas2022high,bianco2022deep}, and integration with other imaging modalities (X-ray \cite{holloway2017savi}, tomography \cite{horstmeyer2016diffraction,zuo2020wide}). Although there are numerous ways to improve the FP technique, like changing optical configurations \cite{pacheco2016reflective,he2018single}, elaborating deep learning \cite{kappeler2017ptychnet,nguyen2018deep,bianco2022deep}, we would like to consider the technique development from an illumination perspective. 

In the history of FP development, researchers have made many efforts to modify illumination strategies, the ultimate goal of which is to improve imaging throughput (resolution, FOV) and speed. To categorize the illumination strategy fundamentally, we derive a universal and relatively fair metric for comparing different imaging techniques. This metric, termed space-frequency-time product (SBP-T \cite{wu2021imaging}), combines space-frequency product (SBP \cite{lohmann_spacebandwidth_1996}) which reflects the throughput of imaging and the acquisition time. With the theoretical analysis of SBP-T, we obtain the principle of designing the illumination strategy in FP, which includes increasing 
 numerical aperture and decreasing low-resolution image acquisition times.

In the following sections, we investigate the illumination strategies for FP performance improvement. We first introduced the basic concept of FP, working principles, and processing pipelines in \autoref{sec:concept}; In \autoref{sec:sbpt} we derive the comprehensive standard SBP-T for the judgment of the microscope's performance. Based on the analysis of SBP-T, we demonstrated three categories of illumination strategies to improve it in \autoref{sec:improve_sbpt}: (1) increase objective numerical aperture (\autoref{sec:objNA}) (2) increase illumination numerical aperture (\autoref{sec:illNA}); (3) reduce requisite sampling time (\autoref{sec:samp-reduction}). Finally, we summarize the paper and discuss the future development of FP in \autoref{sec:summary}.

\section{Principle of Fourier Ptychography}\label{sec:concept}
Fourier Ptychography (FP) is a technique that utilizes ptychography in the Fourier domain. It synthesizes multiple images taken at different spatial frequencies. By post-processing the obtained set of images, an image with a large FOV and high resolution is obtained. Here we briefly introduce the basic architecture and processing principle of FP.

\subsection{Forward imaging model}
\begin{figure}[htbp]
    \centering
    \includegraphics[width=0.5\textwidth]{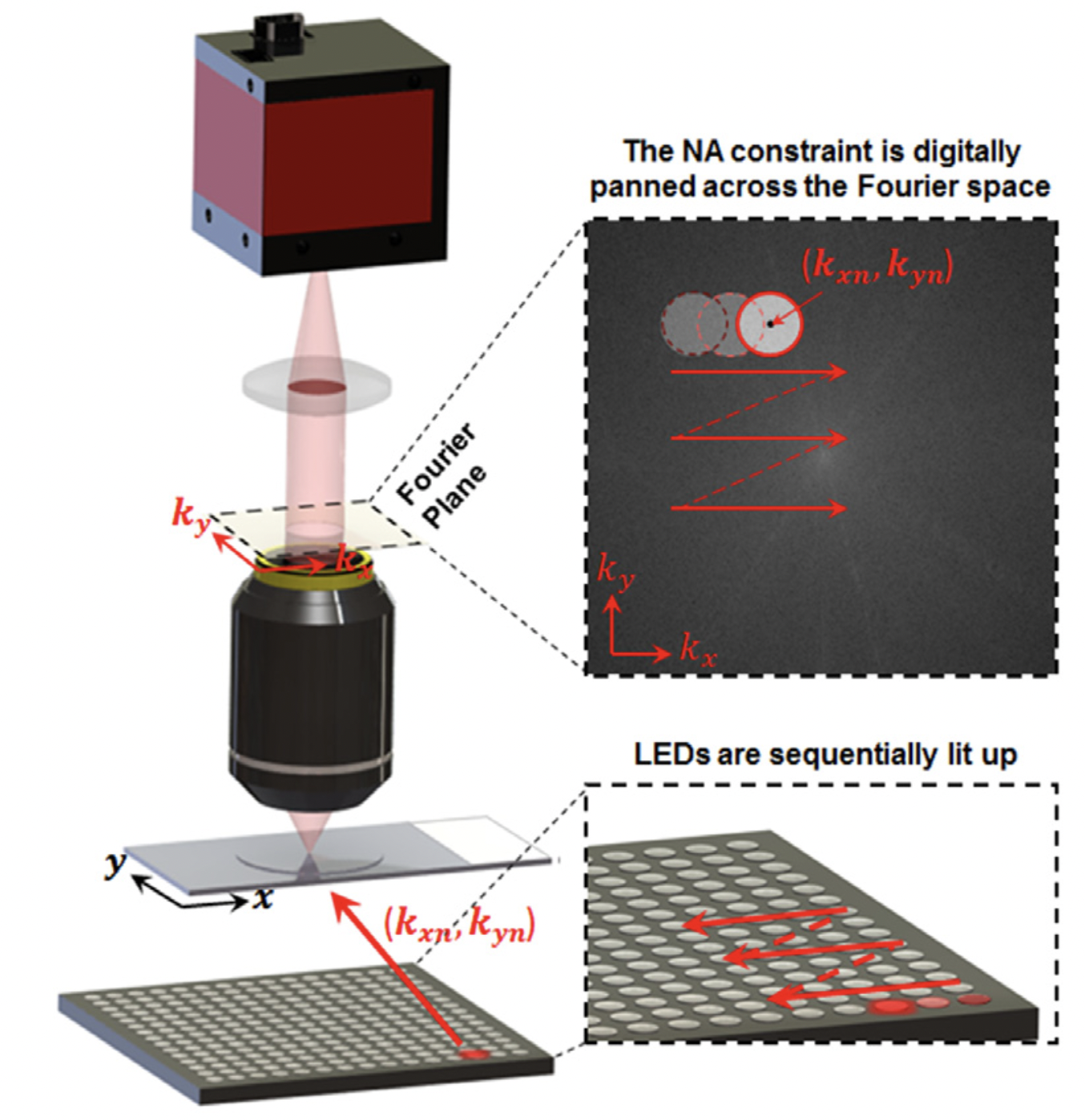}
    \caption{Optical setup of Fourier ptychography. Adapted from Ref. \cite{ou2013quantitative}.}
    \label{fig:set-up}
\end{figure}

The optical setup is shown in \autoref{fig:set-up}. The image sensor and sample are conjugated by a tube lens and an objective. Below the sample is the LED array that will be lit in sequence to provide spatial-frequency-varied illumination.

Assume the sample has the complex amplitude $A(\r),\ \r=(x,y)$. When illuminated with the $m^{th}$ LED light that has spatial frequency $\k_m=(k_{xm,}k_{ym})$, the input amplitude to the imaging system is $E_{obj}(\r)e^{i\k_m\cdot\r}$. The output optical field at the camera plane is 
\begin{equation}
E(\r)=p(\r)*E_{obj}(\r)e^{i\k_m\cdot\r},    
\end{equation}
where $p$ is the system (coherent) point spread function, and $*$ denotes convolution. 

In the Fourier domain, the field is
\begin{equation}
\begin{aligned}
    \hat E(\k)&=P(\k)\cdot \hat E_{obj}(\k-\k_m),\\
    I(\r)&=\left|\mathcal{F}^{-1}\left[\hat E(\k)\right]\right|^2,
\end{aligned}
\end{equation}
where $\hat E_{obj}$ is the output at the objective focal plane, $P$ is the pupil function (or coherent transfer function) of the microscope.

\subsection{Recovery process}
The processing of FP is an image synthesis procedure that iteratively completes the spatial frequency components. It uses captured low-resolution images $I_{lm}$ to synthesize a high-resolution image $I_h$ \cite{zheng2013wide}, where $m\in\{1,2,\dots,N\}$ indicates the image is related to LED light that has the incident angle with spatial frequency $\k_m=(k_{xm},k_{ym})$. 

\subsubsection{Optimization model}
The recovery can be formulated as the following optimization problem:
\begin{equation}
\min _{\hat E_{obj}(\k), P(\k)}
 \sum_{m=1}^{N}
 \norm{I_m(\r)-\left|\mathcal{F}\left[P(\k)\cdot \hat E_{obj}(\k-\k_m)\right]\right|^2}_2.
 \label{eq:recovery-FP}
\end{equation}

To solve the above optimization problem, an iterative phase-retrieval-like algorithm is used. Assume the reconstructed high-resolution image $I_h$ is the intensity of electrical field $E_h=\sqrt{I_h}e^{i\varphi_h}$. Firstly, an initial guess is made by setting $\varphi_h=0$ and $I_{h}^{(0)}=\mathrm{Up}\left(I_{l0}\right)$, where $\mathrm{Up}$ denotes upsampling. Once the initial guess is done, an alternative projection iteration process starts: 
\begin{enumerate}
    \item Take Fourier transform of the last updated electrical field $E_h^{(n)}$: $\hat E_h^{(n)}=\mathcal{F}\left(E_h^{(n-1)}\right)$
    \item Perform the selective spatial frequency pupil filtering: select the right pupil position in the spatial frequency domain, which is moved from the origin to $\k_m=(k_{xm},x_{ym})$ by the angled incident light. We get $\psi^{(n)}(\k)=\mathrm{P}(\k)\cdot \hat E_h^{(n)}(\k-\k_m)$. (Since the pupil function is the size of the low-resolution image, and $E_h$ is the size of the high-resolution image, $E_h$ needs to move its center to match $P$. For a brief, we just use $\mathrm{P}\cdot \hat E_h$ in the following derivation)
    \item Fourier transform back to the image plane: $E_h'^{(n)}=\mathcal{F}^{-1}\left(\psi^{(n)}\right):=\sqrt{I_h'}e^{i\varphi_h'^{(n)}}$, where $\mathrm{P}$ indicates the pupil function.
    \item Utilize the angled light image constraint: replace the amplitude within the pupil region of $E_h'^{(n)}$ with the captured low-resolution image $I_{lm}$. We get $E_h^{(n+1)}=\sqrt{I_{lm}}e^{i\varphi_h'^{(n)}}$.
\end{enumerate}

In one loop, all $\k_m$ of the captured image sequence are used. It can be run repeatedly to get better results, and usually several is enough. Since the recovery process is adapted from the phase retrieval algorithm, the sample phase can be reconstructed, and therefore FP has the ability of quantitative phase imaging \cite{ou2013quantitative}.

\subsubsection{Aberration correction scheme}
The ideal recovery process assumes no aberration, which means pupil function $\mathrm{P}(\k)$ in \autoref{eq:recovery-FP} is ignored. But in experiments, aberrations such as astigmatism and defocus often appear, and it is necessary to correct them. Therefore, aberration correction sometimes becomes an inevitable step in the recovery process.

In the FP literature, there are two correction schemes \cite{zheng2016fourier}: (1) adaptive pupil compensation and (2) joint high-resolution and pupil recovery. Both schemes require an appropriate estimation of the pupil function $\mathrm{P}$. For example, in the defocus correction, $\mathrm{P}$ will add a phase term $\exp\left(ik_z z\right)$ to denote that the focal plane of the objective deviates the sample plane by a distance $z$.

\paragraph{Adaptive pupil compensation}
In the adaptive scheme, pupil estimation follows a closed-loop iteration as adaptive optics \cite{zheng2016fourier}. It searches pupil function (typically the defocus depth $z$) that minimizes the reconstruct metrics (usually the data fidelity measurement of $\|\sqrt{I_{hm}}-\sqrt{I_{lm}}\|$, where $I_{hm}$ is the low-resolution image from the reconstructed high-resolution image that relates to $\k_m$ LED light.

In Ref. \cite{bian2013adaptive}, the algorithm aims at optimizing the term $\sum_{\r}|\sqrt{I_{hm}}-\sqrt{I_{lm}}|$,  To do this, it adds a correction factor $c_m$ \cite{bian2013adaptive}:
\begin{equation}
    c_m=\frac{\sum_{\r}I_{hm}}{\sum_{\r}I_{lm}},
\end{equation}
and change step 4 replacement in the recovery procedure to $E_h=\sqrt{c_mI_{lm}}e^{i\varphi_h'}$. This formula can be considered a weighted update where good reconstruction fidelity of each spatial frequency image takes the higher weight.

\paragraph{Joint pupil recovery} \label{subsubsec:joint-pupil-rec}
In the joint scheme, pupil estimation and high-resolution object recovery are performed simultaneously in the  iteration right after step 4.

The additional update rule is the following \cite{ou2014epfr}:
\begin{equation}
    \begin{aligned}
        E_h^{(n+1)}&=E_h^{(n)}+\alpha\cdot\left(I_{hm}-I_{lm}\right)\frac{P^{\dagger (n)}}{\max\left(\left|P^{(n)}\right|^2\right)},
        \\
        P^{(n+1)}&=P^{(n)}+\beta\cdot\left(I_{hm}-I_{lm}\right)\frac{E_h^{\dagger(n+1)}}{\max\left(\left|E_h^{(n+1)}\right|^2\right)}.
    \end{aligned}
\end{equation}
Here $\dagger$ is the conjugation operator, $\alpha,\beta$ are the step size of the update (can be set to $1$). The update can be seen as compensation for the underestimated pupil \cite{maiden_improved_2009}.

\section{Space-bandwidth-time product} \label{sec:sbpt}
Space-bandwidth product (SBP) is a critical characteristic in optical systems \cite{lohmann_spacebandwidth_1996,park2021review}. It is defined by the range of location and spatial frequencies within which the signal is nonzero. 

Formally, we denote the numerical aperture of the optical as $\NA$, which consists of the objective numerical aperture and $\NA_o$ and illumination numerical aperture $\NA_i$:
\begin{equation}
    \NA=\NA_o+\NA_i,
\end{equation}
and the system resolution $\R$ can be computed as 
\begin{equation}
    \R=\frac{\lambda}{2\NA}.
\end{equation}

Next, we denote the objective lens field-of-view as $\FOV$, which is proportional to the spatial location range. Note that the cutoff spatial frequency is $\nu=1/\R$, then SBP can be computed as \cite{pan2020high}:
\begin{equation}
\begin{aligned}
    \mathrm{SBP}&\sim\FOV\cdot\nu^2=\frac{\FOV}{\R^2}\\
    &=\frac{\lambda^2\cdot \FOV}{4\NA^2}.
\end{aligned}
\end{equation}

The equation of SBP shows its relationship with $\FOV$ and $\NA$. It is worth mentioning that there is a non-linear relation between $\FOV$ and $\NA$. In general, with a high-NA objective lens, there will be a relatively small SBP because the $\FOV$ of an objective lens does not increase at the same pace as that of its $\NA$ \cite{park2021review}.

Image acquisition speed is also essential in many biological applications. Therefore, the space-bandwidth-time product (SBP-T) is considered a more comprehensive standard in microscopy performance comparison \cite{wu2021imaging}. It is defined as the SBP divided by its acquisition time $T$. 

To derive $\text{SBP-T}$, we assume the mean processing time per pixel is $\Delta t$ (including exposure and signal transmission), and the effective total pixels in one capture are estimated as $P=\FOV/\R$. Note that the system performance will decrease if the actual camera pixel $P_{\text{cam}}$ is less than $P$, which can be avoided by changing to a better camera. Let the single-pixel computation time be $\Delta t$, and the necessary number of captures be $N$. The image acquisition time is
\begin{equation}
    T=NP\Delta t=N\frac{\FOV}{\R}\Delta t.
\end{equation}

Therefore, the expression of SBP-T is 
\begin{equation}
\begin{aligned}
    \text{SBP-T}&=\frac{\mathrm{SBP}}{T}=\frac{1}{N\R\Delta t}\\
    &=\frac{2\NA}{N\lambda\Delta t}\\&=\frac{2(\NA_{o}+\NA_{i})}{N\lambda\Delta t}.
    \label{eq:SBP-T}
\end{aligned}
\end{equation}



\begin{figure}[htbp]
    \centering
    \includegraphics[height=0.25\textheight]{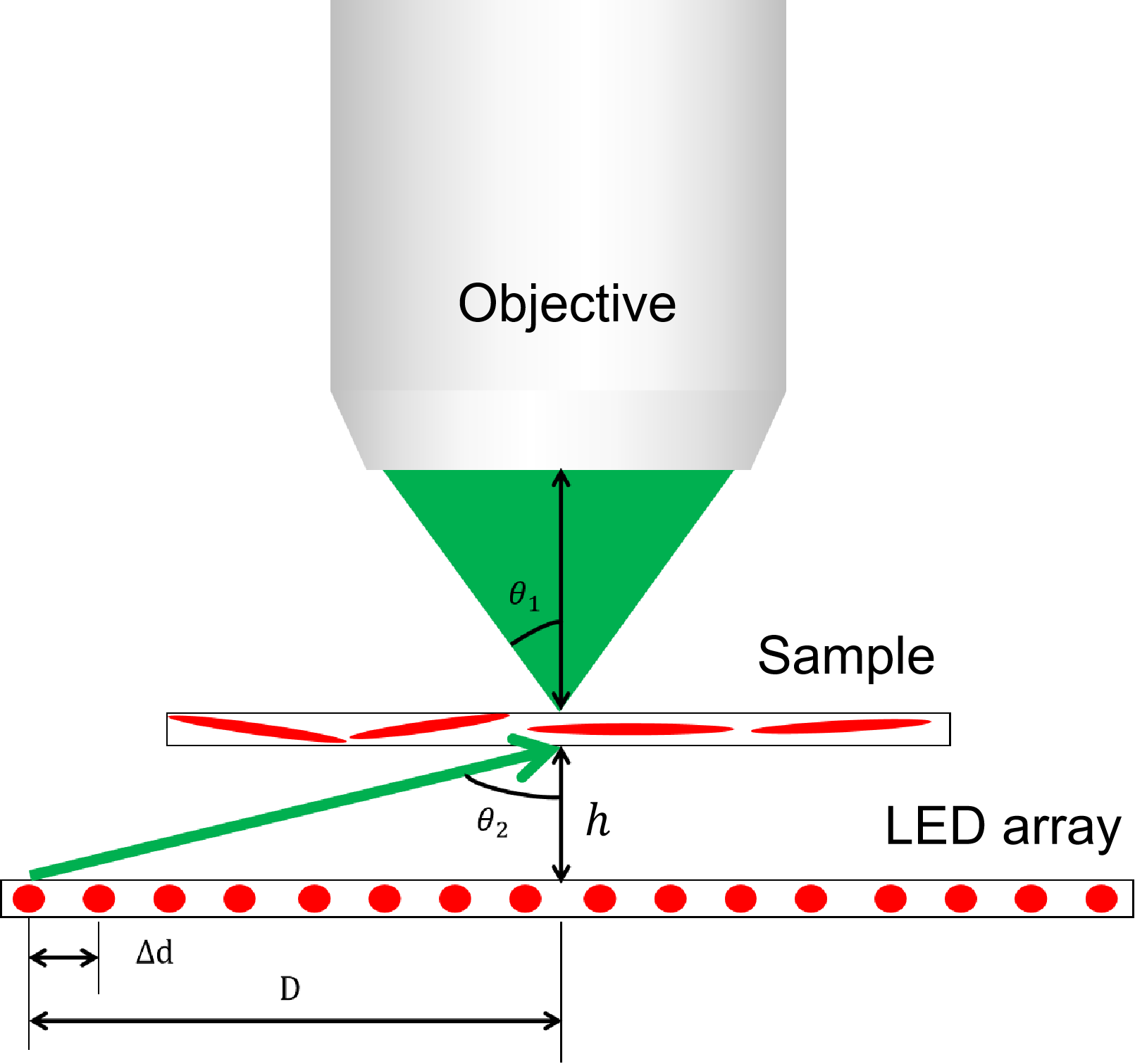}
    \caption{One-dimensional illustration of estimating required acquisition time.}
    \label{fig:sbpt-1}
\end{figure}

Furthermore, if we assume a common FP illumination strategy (i.e., the LED array will be lit one by one), the required acquisition times $N$ can be computed as (\autoref{fig:sbpt-1})
\begin{equation}
\begin{aligned}
    \NA_{i}&=n\sin\theta_2=\frac{nh}{\sqrt{D^2+h^2}}\implies D=\frac{\NA_i h}{\sqrt{n^2-\NA_i^2}},\\
    N&=\left(\frac{2D}{\Delta d}\right)^2=\frac{4h^2}{\Delta d^2\left[(n/\NA_i)^2-1\right]},
\end{aligned}
\end{equation}
where $h$ is the fixed distance between the sample and the LED array, and $D$ is the half length of the LED array, corresponding to the maximum illumination NA. $\Delta d$ is the LED pitch, and $n$ is the refractive index between the LED and sample (illustrated in \autoref{fig:sbpt-1}).

Therefore, the final expression of SBP-T in common LED illumination is
\begin{equation}
    \text{SBP-T}=\frac{\Delta d^2 (\NA_o+\NA_i) \left[(n/\NA_i)^2-1\right]}{2h^2\lambda\Delta t}.
    \label{eq:SBP-T-3}
\end{equation}

\section{Illumination strategies for SBP-T improvement} \label{sec:improve_sbpt}
We have derived $\text{SBP-T}$ for common LED illumination FP (\autoref{eq:SBP-T-3}). For analysis, \autoref{eq:SBP-T} gives us theoretical directions for improving FP. To be more specific, we set the goal as maximizing $\text{SBP-T}$, and one can achieve this by three means: increasing $\NA_{o}$, increasing $\NA_{i}$, or decreasing required acquisition time $N$. Besides, we do not discuss shortening the exposure and pixel transferring time $\Delta t$ here, which can also improve SBP-T but is straightforward by increasing light transport efficiency.

\subsection{Increasing illumination NA} \label{sec:illNA}
There are three approaches proposed to increasing $\NA_i$: (1) immersing in the high refractive-index medium; (2) using a large-angle illumination condenser; (3) using a reflection-mode configuration.

\begin{figure}[htbp]
    \centering
    \includegraphics[width=0.9\textwidth]{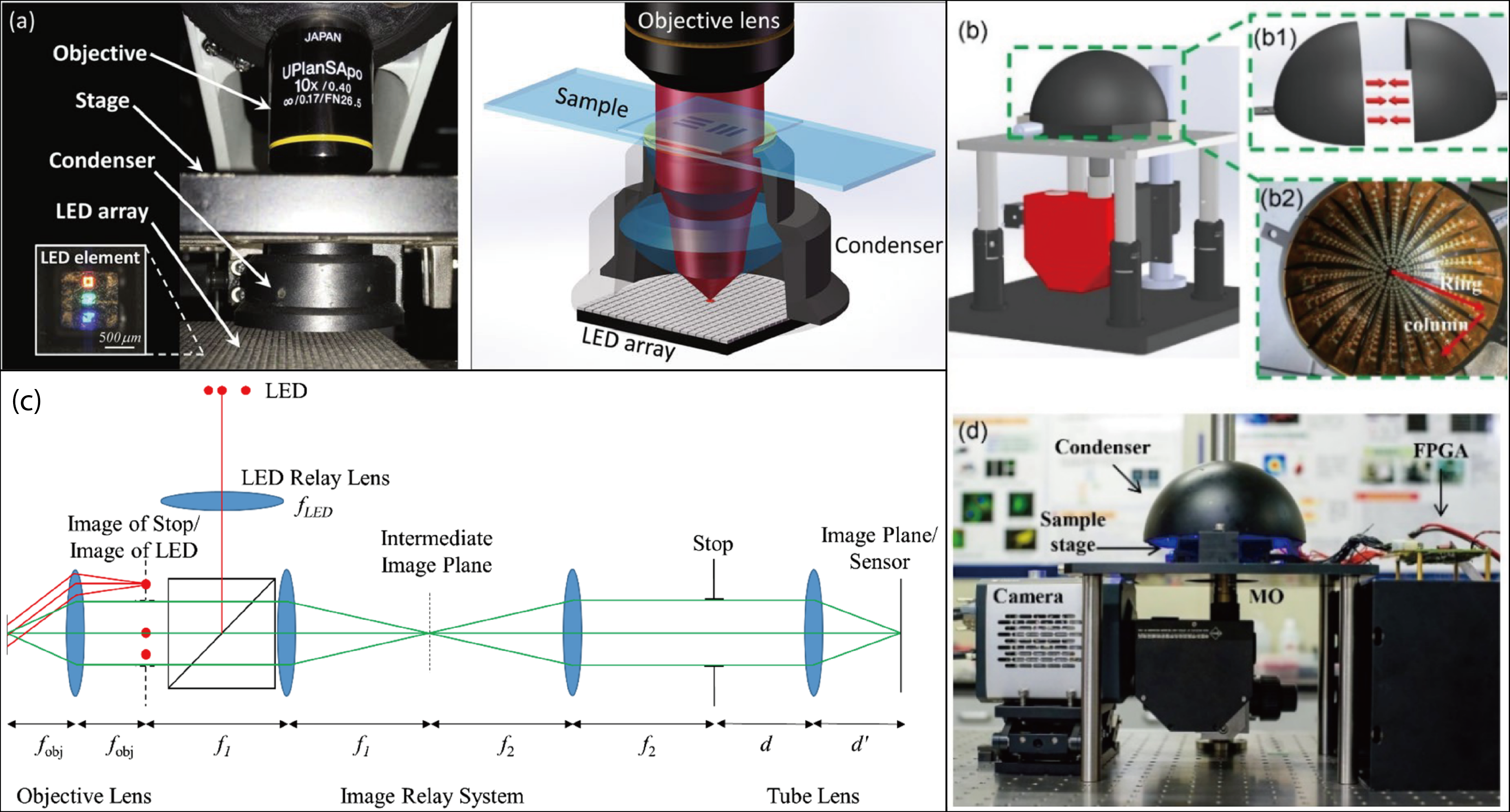}
    \caption{Three approaches for illumination NA improvement. (a) immersing condenser in oil \cite{sun2017resolution}. (b,d) using hemisphere condenser lens \cite{pan2018subwavelength}. (c) reflective FP setup \cite{pacheco2016reflective}.}
    \label{fig:ill-na}
\end{figure}

As shown in \autoref{fig:ill-na} (a), Sun \textit{et al.} straightforwardly designed an oil-immersion condenser LED array to improve $ \NA_{i}$ \cite{sun2017resolution}. The high refractive-index oil enables the condenser to produce higher spatial frequency incident light, which endows them to achieve a final effective NA equal to 1.6, with a 10x, 0.4NA objective lens. In \autoref{fig:ill-na} (b,d), Pan \textit{et al.} designed a hemisphere digital condenser to provide high-angle programmable plane-wave illuminations of 0.95NA and realized a subwavelength resolution (a half-pitch resolution of 244 nm with the incident wavelength of 465 nm across a wide FOV of 14.60 $\text{mm}^2$).

Pacheco \textit{et al.} demonstrated a reflective FP architecture in Ref. \cite{pacheco2016reflective}, and the optical setup is shown in \autoref{fig:ill-na} (c). By placing the LED at the back focal plane of the objective and setting the position of the LED to be greater than the radius of the objective’s entrance pupil, the illumination NA will be equal to the objective NA, hence the maximum synthesized NA is $2\NA_o$. Furthermore, by placing the imaging stop outside the illumination path in the reflective FP, the NA of the illumination can be much greater than the NA of the imaging system. They experimentally demonstrated a synthesized NA increased by a factor of 4.5 using the proposed optical concept.

\subsection{Increasing objective NA} \label{sec:objNA}
Frequency mixing between the object and the illumination allows the recovery of object frequencies beyond the diffraction-limited detection bandpass. by using nonzero-spatial-frequency illumination, the high frequency of the object beyond the detection bandpass of the objective lens is encoded (or moved) into a lower frequency and hence can be captured. This is equivalent to the increase of effective objective NA. As demonstrated in blind structured illumination microscopy \cite{mudry2012strucyured}, using various speckle illumination, it is possible to obtain better resolution (larger $\NA_{o\eff}$) without knowing the exact illumination pattern. This illumination strategy is also transplanted in FP architecture, which we term as \textit{speckle illumination}. 

\begin{figure}[htbp]
    \centering
    \includegraphics[width=0.95\textwidth]{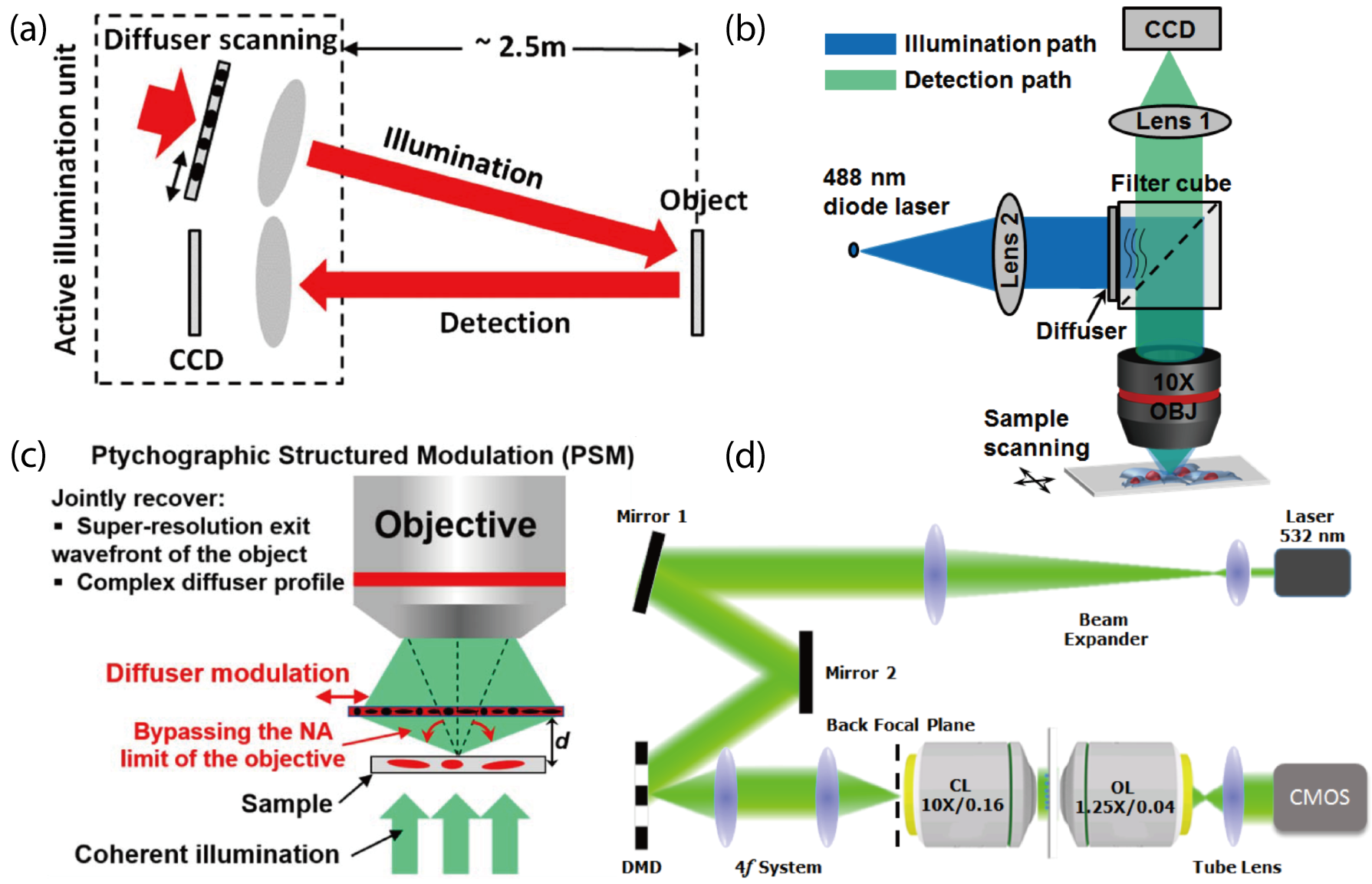}
    \caption{Speckle illumination methods for $\NA_{o\eff}$ improvement. (a) speckle illumination via diffuser scanning \cite{dong2015incoherent}.  (b) speckle illumination via sample scanning. (c) speckle illumination with a diffuser placed behind the sample \cite{song2019super}. (c) speckle illumination using a digital micromirror device (DMD) \cite{kuang2015digital}.}
    \label{fig:speckle-ill}
\end{figure}

When speckle illumination was first proposed in FP \cite{dong2014high}, it substituted the traditional LED array with a laser and diffuser. The semitransparent diffuser was placed before the sample, and the sample was translated to obtain various speckle patterns (\autoref{fig:speckle-ill} (b)). This setup was modified by making the diffuser the moving part \cite{dong2015incoherent} (\autoref{fig:speckle-ill} (a)). In Ref. \cite{song2019super}, Song \textit{et al.} placed the diffuser behind the sample (\autoref{fig:speckle-ill} (c)). Benefiting from the novel setup, the sample thickness becomes irrelevant, and they can focus on the wavefront that exits the sample. They showed a 4.5-fold resolution gain over the diffraction limit. 

\begin{figure}[htbp]
    \centering
    \includegraphics[width=0.7\textwidth]{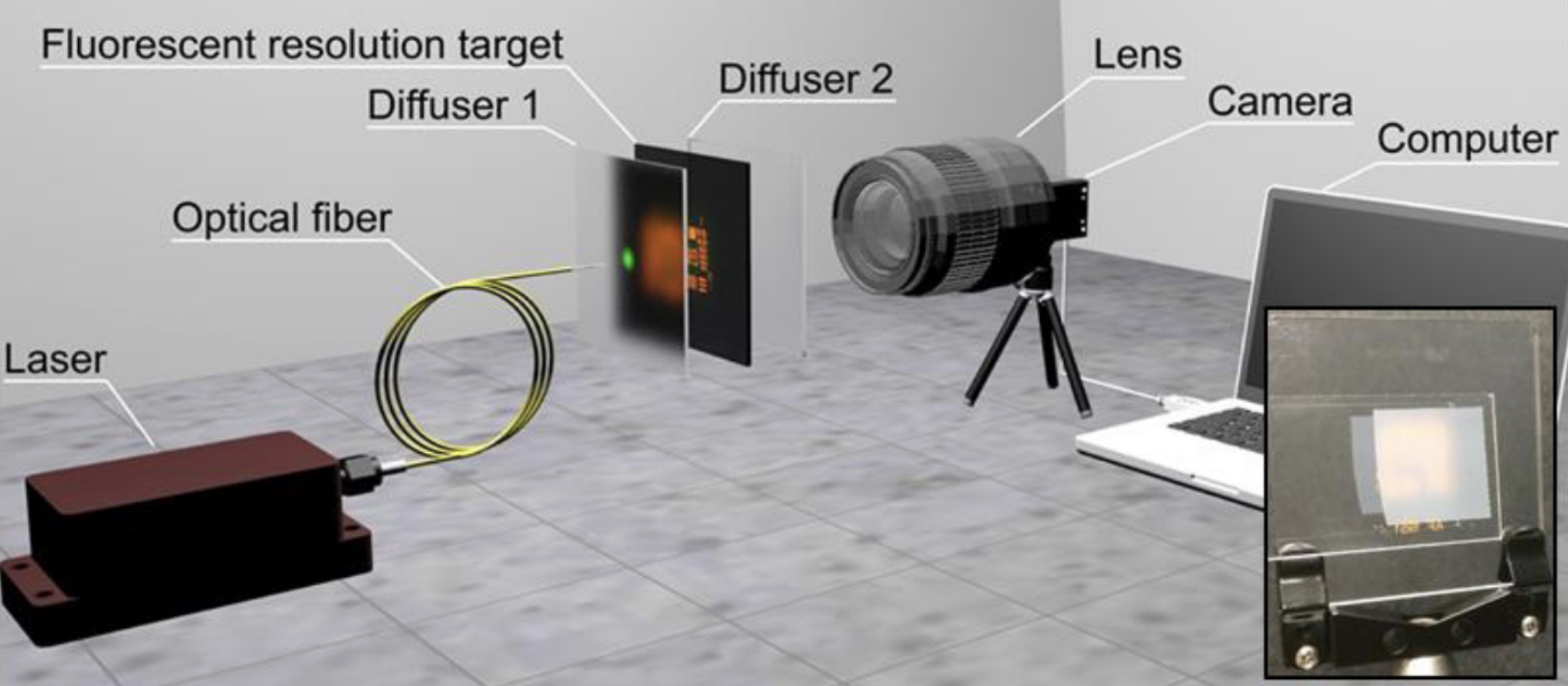}
    \caption{The system setup of the reported double diffuser strategy for speckle illumination FP. Adapted from Ref. \cite{guo2018-13fold}.}
    \label{fig:double-diffuser}
\end{figure}

In 2018, Kaikai Guo \textit{et al.} further developed the diffuser speckle illumination method \cite{guo2018-13fold}. They obtained 13 times resolution enhancement by sandwiching the sample between two diffusers (\autoref{fig:double-diffuser}). By tilting the input wavefront, they raster scan the unknown speckle pattern and capture the corresponding low-resolution image that passes through the turbidity layer. The front-behind diffusers, together with the joint object-speckle-PSF reconstruction algorithm allow them to grasp higher object frequency that is otherwise untouchable.
 
A new approach for speckle generating is proposed by elaborating on the digital micromirror device (DMD) \cite{kuang2015digital}. \autoref{fig:speckle-ill} (d) shows the configuration of the DMD-based laser-illumination FPM system. The use of DMD can significantly accelerate the illumination pattern generation speed and get rid of the irritating procedure of moving the sample or diffuser.

\begin{figure}[htbp]
    \centering
    \includegraphics[width=0.55\textwidth]{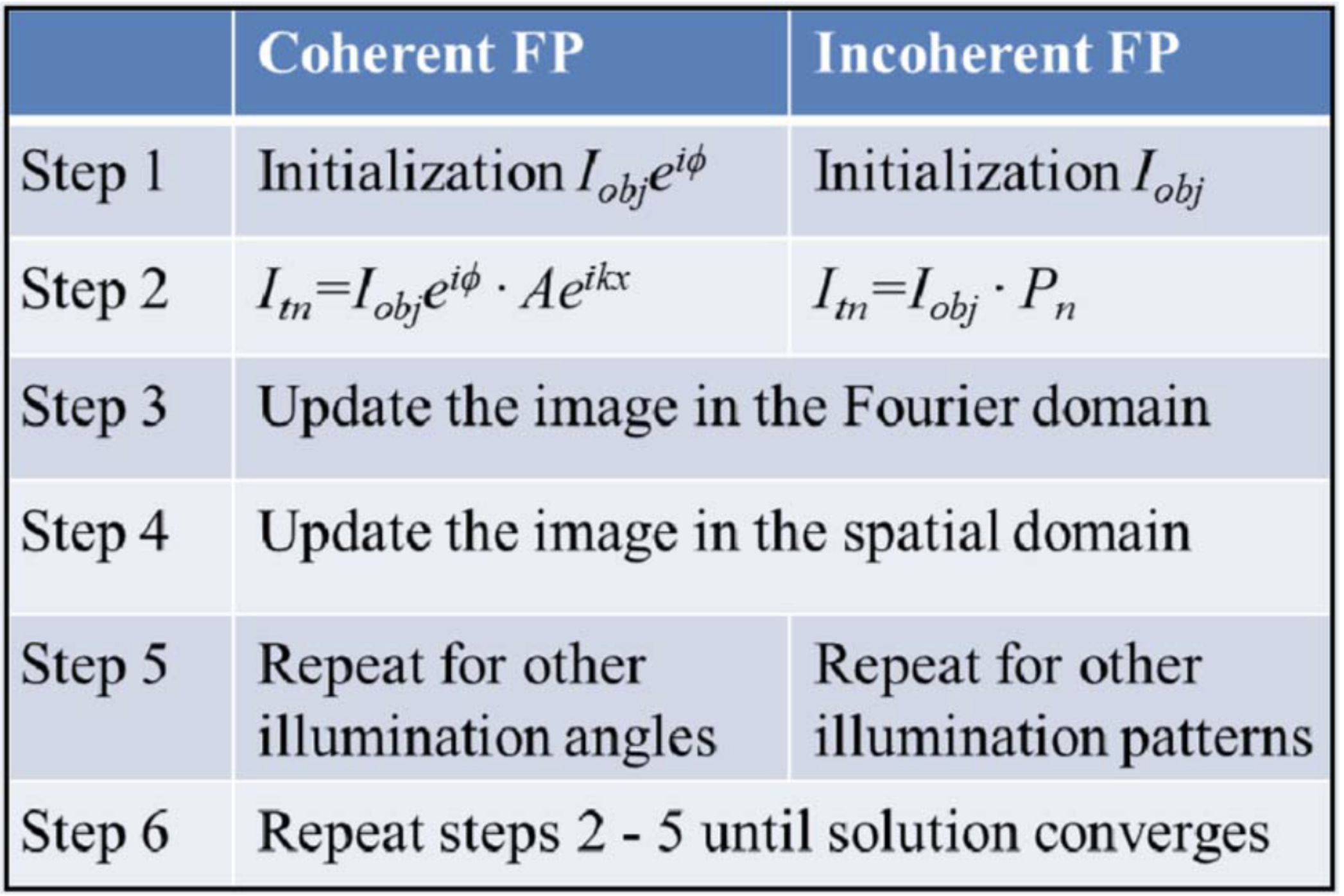}
    \caption{Comparison of LED-array illumination FP and speckle illumination FP. Adapted from Ref. \cite{dong2015incoherent}.}
    \label{fig:cmp-speckle-ill-algo.png}
\end{figure}

The reconstruction process is similar to the original FP recovery, where the pupil function becomes the unknown speckle pattern, and joint pupil recovery (\autoref{subsubsec:joint-pupil-rec}) for speckle pattern guessing is always employed. \autoref{fig:cmp-speckle-ill-algo.png} gives a direct overview of reconstruction algorithms in speckle-illumination FP.

As an extension of speckle illumination FP, a method called near-field Fourier ptychography was proposed for photography beyond microscopy. This is realized by placing the object at a short defocus distance with a large Fresnel number, and a speckle pattern with fine spatial features is cast on the object. The difference between near-field FP and original speckle illumination FP is the propagation of diffused light. A diffraction propagation equation has to be inserted into the reconstruction process.
 
\subsection{Acquisition time reduction} \label{sec:samp-reduction}
\subsubsection{Sparse illumination}
To address the efficiency problem in current FPM, some researchers investigated the spatial spectrum redundancy of natural images in the Fourier domain and proposed sampling strategies to decrease the requisite illumination angles during acquisition and correspondingly decrease the shutter count $N$ of the FP technique.


\begin{figure}[htbp]
    \centering
    \includegraphics[width=0.85\textwidth]{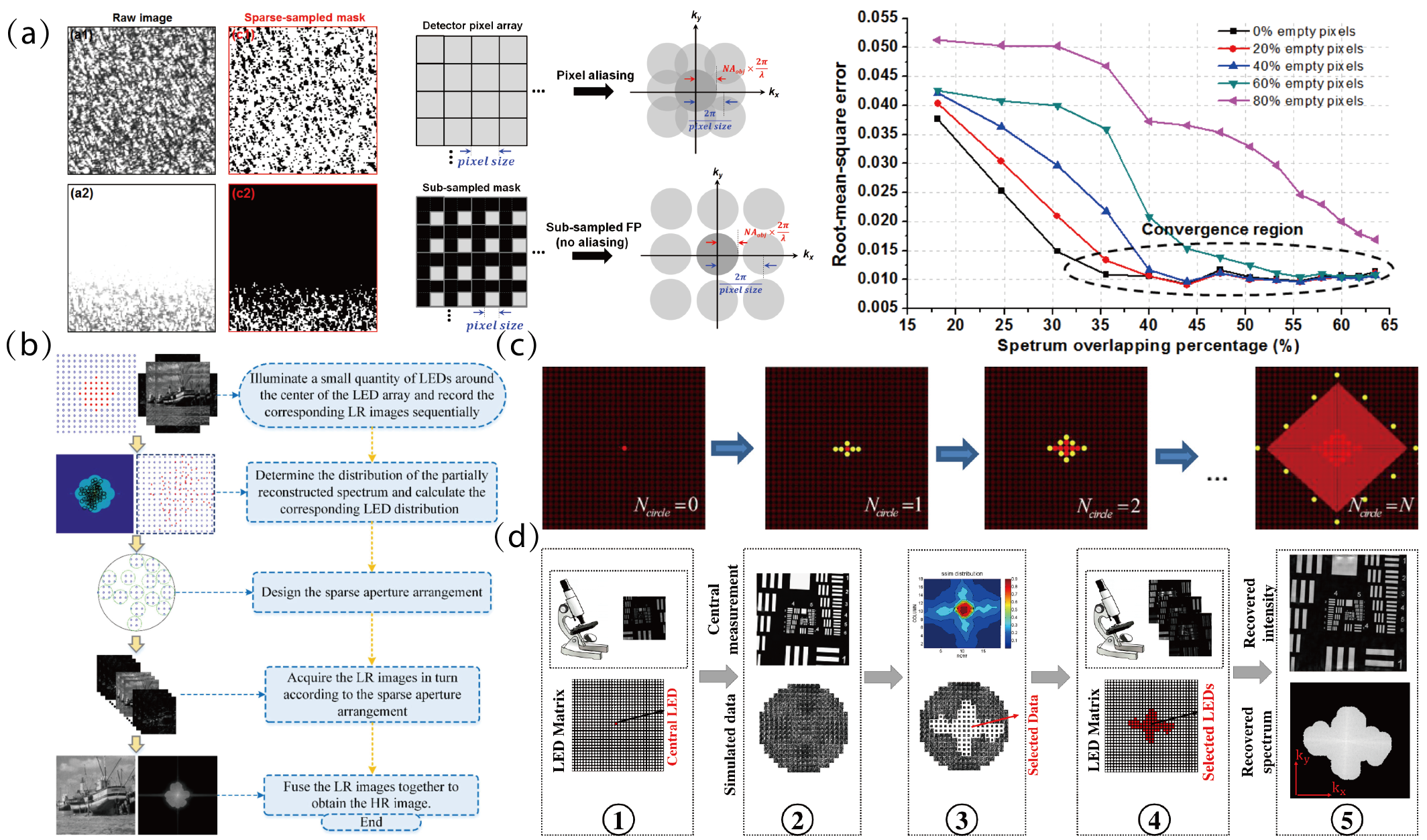}
    \caption{Sparse illumination for FP acquisition reduction. (a) spatial sparse sampling via masks, spectral sparse sampling via subsampled pixels, spatial-spectral redundancy curve \cite{dong2014sparsely}. (b) Spectral-distribution-based illumination LED selection \cite{cheng2019sparse}. (c) Rhombus-shaped LED array illumination \cite{mao2020efficient}. (d) SSIM-based spectral selection scheme \cite{zhang2015self}.}
    \label{fig:sparse-ill}
\end{figure}

In 2014, Dong \textit{et al.} investigated the spectral-spatial data redundancy requirement in FP, which is the theoretical basis of the sparse illumination strategy \cite{dong2014sparsely}. They experimentally examined how much spectrum overlapping (in the spectral domain) and how many pixels need to update is needed (in the spatial domain) for a decent FP reconstruction. Finally, they proposed a sparse update using a mask in the spatial domain (which essentially drops the over-exposure and under-exposure pixels) and a sub-sampled detector pixel array. By using the proposed sparse sampling framework, they shorten the acquisition time of the FP by $\sim 50\%$.
 
To achieve sparse sampling, Bian \cite{bian2014content} proposed a highly efficient method termed adaptive Fourier ptychography (AFP) in 2014. It applies content-adaptive illumination for FP to capture the most informative parts of the scene’s spatial spectrum. By selectively illuminating circle by circle towards the high-frequency direction, it reduces acquisition time by around 30\%–60\%. In 2015, Zhang \cite{zhang2015self} proposed a self-learning-based method, which utilizes the single captured central-illumination measurement and simulates the angled illumination. Based on the structural similarity metric (SSIM) between the simulated image spectrum and the measurement spectrum, they can select images used in the FPM reconstruction. The number of acquired images can be reduced from 225 to 70, with a time reduction of $\sim 70\%$. The flowchart of the method is shown in \autoref{fig:sparse-ill} (d).
Similarly, Cheng \cite{cheng2019sparse} proposed a sparse illumination method based on maximum object spectral information. The flow chart of their proposed method is shown in \autoref{fig:sparse-ill} (b). Based on the relationship between the spectrum distribution of the specimen and the LED array sampling pattern, they analyze the spectrum of captured center-illumination low-resolution image to manually select the informative position and design a sparse aperture arrangement accordingly. The total number of LEDs sampled is reduced by 40\% to 55\% and the total processing time for the sampling is reduced from 304.31 s to 124.86 s compared with the traditional FPM sampling method. 

\begin{figure}[htbp]
    \centering
    \includegraphics[width=0.55\textwidth]{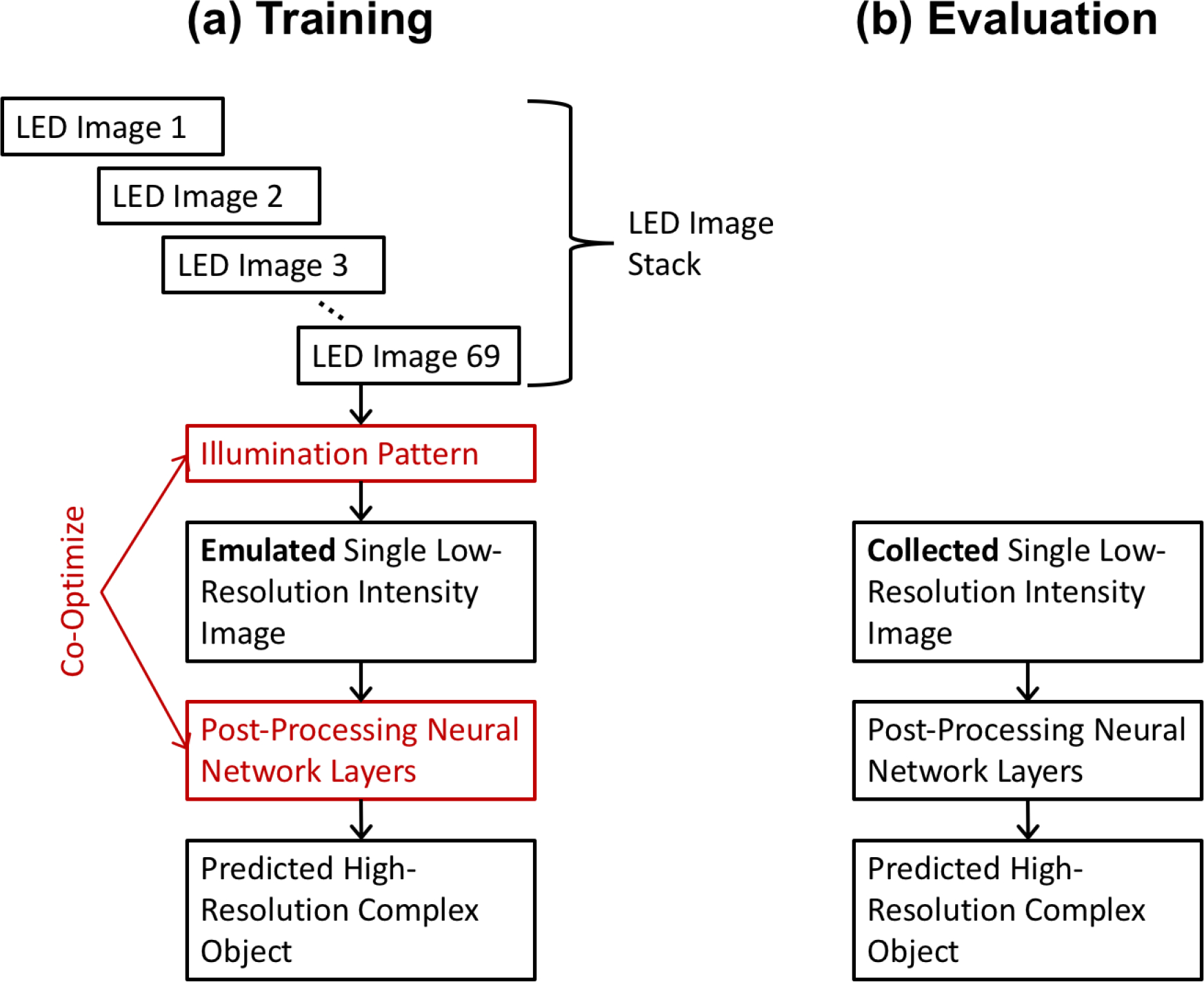}
    \caption{Overview of the training and evaluation steps in optimization. Adapted from Ref.
    \cite{cheng2019illumination}}
    \label{fig:cheng2}
\end{figure}

Great development in deep learning also allows researchers to design illumination patterns intelligently. In 2019, Cheng \cite{cheng2019illumination} used a deep neural network to achieve single-shot imaging, reducing the acquisition time in Fourier ptychographic microscopy by a factor of 69. They proposed an illumination pattern co-optimization framework. In the training phase, shown in \autoref{fig:cheng2} (a), they use deep learning to find a single LED illumination pattern, where the LED pattern corresponds to a neural network layer. In the evaluation phase (shown in \autoref{fig:cheng2} (b)), they first program the LED matrix with a static pattern. Then, for each sample, they take a single image and process it through the trained neural network layers to get the high-resolution complex object.

As a kind of sparse illumination, Mao \cite{mao2020efficient} proposed a method based on the rhombus-shaped LED array. They selected the most important LEDs as the rhombus circle expanded. Compared to the reconstruction process under conventional illumination, it is reduced the total time from 99.126 s to 25.396 s, with a reduction of $\sim 77\%$. The selected N LEDs by following the shape of the rhombus are shown in \autoref{fig:sparse-ill} (c).



\subsubsection{Multiplexing}

\begin{figure}[htbp]
    \centering
    \includegraphics[width=0.8\textwidth]{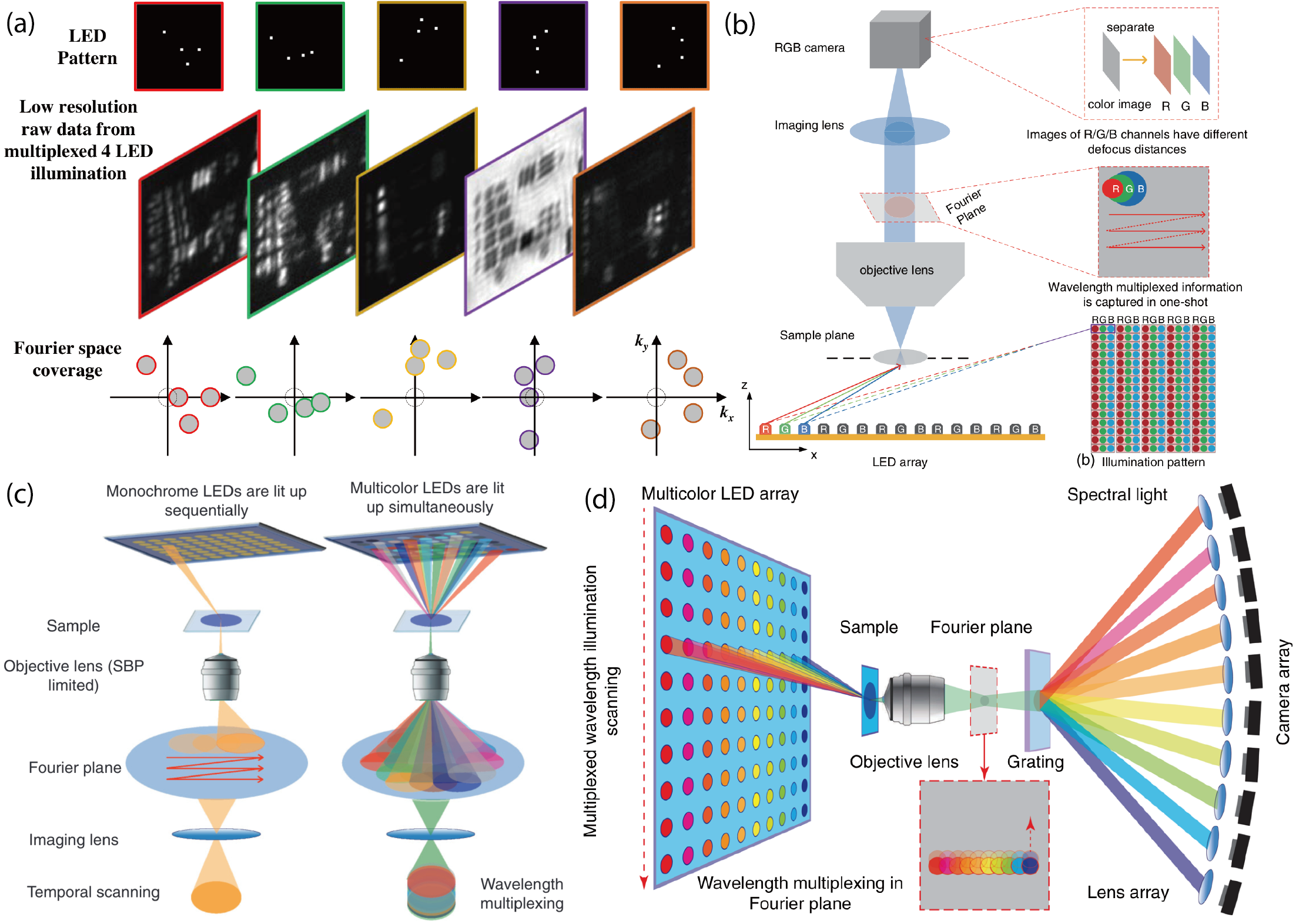}
    \caption{Multiplexed illumination. (a) Angle-multiplexing \cite{tian2014multiplexed}. (b) RGB wavelength-multiplexing demonstrated in Ref. \cite{zhou2017fourier}. (c) Concept of wavelength multiplexing with multicolor LEDs \cite{zhou2017fourier}. (d) Concept of wavelength multiplexing with grating and camera array for accelerated detection \cite{zhou2017fourier}.}
    \label{fig:multiplexing}
\end{figure}

Multiplexing is also an effective acceleration approach in FP imaging. In 2014, Tian \textit{et al.} proposed an angle-multiplexing illumination (also called source-coded illumination \cite{tian_computational_2015}), where multiple randomly selected LEDs are used to illuminate each image \cite{tian2014multiplexed}. For example, illumination patterns designed according to random codes and resulting images are shown in \autoref{fig:multiplexing} (a). With an appropriate reconstruction algorithm, this scheme allows them to fill the Fourier space faster, which can greatly reduce the sampling time and data size in FP. They demonstrated the large SBP-T advantages of angle-multiplexing illumination (46 megapixels/s) with improved initialization strategy and reconstruction algorithm in Ref. \cite{tian_computational_2015}.

In contrast to angle-multiplexing for faster filling in the Fourier domain, another multiplexing strategy is wavelength multiplexing. The proposed wavelength multiplexing approach in Ref. \cite{zhou2017fourier} can speed up the acquisition process of FP several folds. \autoref{fig:multiplexing} (c) illustrates the concept of multicolor LEDs lighting up simultaneously to illuminate different circular regions in a sample’s Fourier plane. In this way, different wavelengths are used to label the different spatial frequencies of the samples, which can be decoded via the corresponding algorithms. In experiments, they demonstrated an RGB multiplexed illumination and used an RGB camera for channel de-multiplexing (\autoref{fig:multiplexing} (b)). They also proposed an accelerated detection approach by hardware de-multiplexing, using a grating and camera array, as shown in \autoref{fig:multiplexing} (d).

Reconstruction algorithms for de-multiplexing is essential to multiplexing methods. Several algorithms based on the multiplexing principle have been proposed simultaneously as the multiplexing strategy is proposed \cite{dong2014spectral,tian_computational_2015,zhou2017fourier}. Algorithm robustness \cite{yeh2015experimental} and acceleration \cite{bostan2018accelerated} have also been investigated. Since the reconstruction algorithm is the post-processing step in FP and has little contribution in improving SBP-T (note that $\Delta t$ in \autoref{eq:SBP-T} is only related to exposure and data transfer time when capturing), we would not give a detailed discussion here.


\section{Summary} \label{sec:summary}
In this paper, we reviewed illumination strategies for improving SBP-T in FP. Based on the analysis of SBP-T, we choose to investigate FP illumination methods from the aspect of NA improvement and necessary image acquisition reduction: (i) For NA improvement, we first discussed methods regarding illumination NA: the utilization of a specially designed condenser lens and a reflective optical setup. Then we compared different speckle illumination configurations for improving objective NA. The speckle illumination in FP borrows the concept of blind-SIM and shows good results in improving the synthetic NA. It is worth mentioning that it endows FP with the ability of fluorescence imaging, as the intensity-varied speckle can encode fluorescence emitting while angle-varied LED illumination cannot. (ii) For image acquisition reduction, we discussed sparse illumination and multiplexed illumination. Both methods demonstrate the ability to reduce the capturing times. There is also a method called state-multiplexing, which uses multiple light sources to illuminate simultaneously the incoherent superposition of the transmission profiles with different coherent states \cite{dong2014spectral}. Since this multiplexing method cannot reduce the acquisition time $N$ effectively, we do not put it in the multiplexing illumination class. 

In addition, we think it is necessary to calculate the SBP-T of each proposed FP method and compare them quantitatively, but we cannot complete that here due to time and device limitations (here we give a brief comparison table with a limited number of methods).

\begin{table}[htbp]
\caption{FP technique comparison without SBP-T.}
\begin{tabular}{|c|c|c|c|}
\hline
Reference                             & Full-pitch Resolution \cite{horstmeyer2016standardizing} & FOV                    & Speed \\ \hline
\textit{Opt. Express} \textbf{30}, 29189 (2022).        & 0.3 um                & $0.65mm\times0.65mm$   & 3s                                  \\ \hline
\textit{Opt. Express} \textbf{27}, 644–656 (2019).      & \textless{}2 um       & /                      & 1.14s                               \\ \hline
\textit{Opt. Express}, \textbf{26}, 23119–23131 (2018). & 0.488  um             & $14.6\ mm^2$           & 4.15s                               \\ \hline
\textit{Sci Rep}\textbf{ 7}, 1187   (2017).             & 308 um                & $2.34\ mm^2$           & /                                   \\ \hline
\textit{J. Biomed. Opt} \textbf{22}, 066006 (2017).     & 1.56 um               & /                      & $\sim$ 100s                         \\ \hline
\textit{Optica},\textbf{ 2},   904–911 (2015).          & \textless{}2 um       & /                      & 0.8s                                \\ \hline
\textit{Opt. Express} \textbf{22}, 20856 (2014).        & 0.66 um               & $15\ um \times 15\ um$ & /                                   \\ \hline
\textit{Nature Photon}\textbf{ 7}, 739–745 (2013).      & 1.56 um, axial 0.3 mm & $120\ mm^2$            & 3min                                \\ \hline
\end{tabular}
\end{table}

Furthermore, we would like to mention the illumination strategy of FP can be designed specifically for different applications. For instance, because the reconstruction algorithm of FP is phase-retrieval-like, it can recover the phase information with good quality. This ability enables FP to be applied in quantitative phase imaging (QPI), and corresponding illumination strategies can be specifically modified. In 2018, a reduced ring illumination was proposed by Sun \textit{et al.} \cite{sun2018high} to reduce the acquisition time in QPI greatly. They analyzed the quantitative phase imaging process in FP and tuned the FP illumination pattern for phase imaging applications. They demonstrated that low-frequency phase components could be completely covered at annular illumination when coherent parameter $S=\NA_i/\NA_o=1$. A video-rate FP based on annular illuminations named AIFPM is reported, which uses an annular illumination scheme by only lighting up LED elements located on a ring with the illumination NA matching the $\NA_{o}$. AIFPM reduces the exposure time for each image to only 10 ms, thus the total data acquisition time is 0.12 s for 12-LED scanning and can be reduced to 0.04 s if the fastest 4-LED scanning mode is used. The results show an SBP-T of 411 megapixels/s for in-vitro Hela cell mitosis and apoptosis at a frame rate 25Hz. Multiplexing has been applied in QPI applications. A single-shot QPI method based on color-multiplexed FPM was proposed by Sun \textit{et al.} \cite{sun2018single}, where three monochromatic intensity images are sufficient to achieve high-accuracy phase retrieval by combining phase deconvolution and FPM iterative refinement.

For future development of FP, we think reflective FP is a significant direction for in-vivo applications. Researchers have demonstrated it \cite{pacheco2016reflective} high illumination NA property, and more improvements are expected. In addition, combining different illumination strategies is also promising. Since speckle illumination has shown a fluorescence imaging ability \cite{dong2014high}, it can be combined with reflective configuration hopefully. Sparse illumination and wavelength multiplexing can also be combined to boost image-capturing speed \cite{zhou2017fourier}. Finally, novel optical setups that extend the FP scheme is also expected. For example, a concept of single-shot FP, with a different optical setup is proposed \cite{he2018single}, encouraging us of the great potential of FP future development.


\paragraph{Acknowledgments} 
We thank Prof. Dawei Di, Prof. Pan Wang, and Prof. Ke Si for their generous teaching in academic writing.

\paragraph{Conflict of interest} 
The authors declare no conflicts of interest related to this article.

\bibliographystyle{unsrt}  
\bibliography{main.bbl}

\begin{thebibliography}{10}

\bibitem{konda2020fourier}
Pavan~Chandra Konda, Lars Loetgering, Kevin~C Zhou, Shiqi Xu, Andrew~R Harvey,
  and Roarke Horstmeyer.
\newblock Fourier ptychography: current applications and future promises.
\newblock {\em Optics express}, 28(7):9603--9630, 2020.

\bibitem{zheng2013wide}
Guoan Zheng, Roarke Horstmeyer, and Changhuei Yang.
\newblock Wide-field, high-resolution fourier ptychographic microscopy.
\newblock {\em Nature photonics}, 7(9):739--745, 2013.

\bibitem{ou2015high}
Xiaoze Ou, Roarke Horstmeyer, Guoan Zheng, and Changhuei Yang.
\newblock High numerical aperture fourier ptychography: principle,
  implementation and characterization.
\newblock {\em Optics express}, 23(3):3472--3491, 2015.

\bibitem{pacheco2016reflective}
Shaun Pacheco, Guoan Zheng, and Rongguang Liang.
\newblock Reflective fourier ptychography.
\newblock {\em Journal of biomedical optics}, 21(2):026010, 2016.

\bibitem{sun2017resolution}
Jiasong Sun, Chao Zuo, Liang Zhang, and Qian Chen.
\newblock Resolution-enhanced fourier ptychographic microscopy based on
  high-numerical-aperture illuminations.
\newblock {\em Scientific reports}, 7(1):1--11, 2017.

\bibitem{pan2018subwavelength}
An~Pan, Yan Zhang, Kai Wen, Meiling Zhou, Junwei Min, Ming Lei, and Baoli Yao.
\newblock Subwavelength resolution fourier ptychography with hemispherical
  digital condensers.
\newblock {\em Optics Express}, 26(18):23119--23131, 2018.

\bibitem{zhang2019near}
He~Zhang, Shaowei Jiang, Jun Liao, Junjing Deng, Jian Liu, Yongbing Zhang, and
  Guoan Zheng.
\newblock Near-field fourier ptychography: super-resolution phase retrieval via
  speckle illumination.
\newblock {\em Optics express}, 27(5):7498--7512, 2019.

\bibitem{dong2014sparsely}
Siyuan Dong, Zichao Bian, Radhika Shiradkar, and Guoan Zheng.
\newblock Sparsely sampled fourier ptychography.
\newblock {\em Optics express}, 22(5):5455--5464, 2014.

\bibitem{kuang2015digital}
Cuifang Kuang, Ye~Ma, Renjie Zhou, Justin Lee, George Barbastathis,
  Ramachandra~R Dasari, Zahid Yaqoob, and Peter~TC So.
\newblock Digital micromirror device-based laser-illumination fourier
  ptychographic microscopy.
\newblock {\em Optics express}, 23(21):26999--27010, 2015.

\bibitem{kappeler2017ptychnet}
Armin Kappeler, Sushobhan Ghosh, Jason Holloway, Oliver Cossairt, and Aggelos
  Katsaggelos.
\newblock Ptychnet: Cnn based fourier ptychography.
\newblock In {\em 2017 IEEE International Conference on Image Processing
  (ICIP)}, pages 1712--1716. IEEE, 2017.

\bibitem{zhou2017fourier}
You Zhou, Jiamin Wu, Zichao Bian, Jinli Suo, Guoan Zheng, and Qionghai Dai.
\newblock Fourier ptychographic microscopy using wavelength multiplexing.
\newblock {\em Journal of biomedical optics}, 22(6):066006, 2017.

\bibitem{sun2018single}
Jiasong Sun, Qian Chen, Jialin Zhang, Yao Fan, and Chao Zuo.
\newblock Single-shot quantitative phase microscopy based on color-multiplexed
  fourier ptychography.
\newblock {\em Optics letters}, 43(14):3365--3368, 2018.

\bibitem{he2018single}
Xiaoliang He, Cheng Liu, and Jianqiang Zhu.
\newblock Single-shot aperture-scanning fourier ptychography.
\newblock {\em Optics Express}, 26(22):28187--28196, 2018.

\bibitem{nguyen2018deep}
Thanh Nguyen, Yujia Xue, Yunzhe Li, Lei Tian, and George Nehmetallah.
\newblock Deep learning approach for fourier ptychography microscopy.
\newblock {\em Optics express}, 26(20):26470--26484, 2018.

\bibitem{cheng2019illumination}
Yi~Fei Cheng, Megan Strachan, Zachary Weiss, Moniher Deb, Dawn Carone, and
  Vidya Ganapati.
\newblock Illumination pattern design with deep learning for single-shot
  fourier ptychographic microscopy.
\newblock {\em Optics express}, 27(2):644--656, 2019.

\bibitem{zhang2019fourier}
Jizhou Zhang, Tingfa Xu, Ziyi Shen, Yifan Qiao, and Yizhou Zhang.
\newblock Fourier ptychographic microscopy reconstruction with multiscale deep
  residual network.
\newblock {\em Optics Express}, 27(6):8612--8625, 2019.

\bibitem{sun2018high}
Jiasong Sun, Chao Zuo, Jialin Zhang, Yao Fan, and Qian Chen.
\newblock High-speed fourier ptychographic microscopy based on programmable
  annular illuminations.
\newblock {\em Scientific reports}, 8(1):1--12, 2018.

\bibitem{xiao2021high}
Yi~Xiao, Shiyuan Wei, Shaolong Xue, Cuifang Kuang, Anli Yang, Maoliang Wei,
  Hongtao Lin, and Renjie Zhou.
\newblock High-speed fourier ptychographic microscopy for quantitative phase
  imaging.
\newblock {\em Optics Letters}, 46(19):4785--4788, 2021.

\bibitem{aidukas2022high}
Tomas Aidukas, Pavan~C Konda, and Andrew~R Harvey.
\newblock High-speed multi-objective fourier ptychographic microscopy.
\newblock {\em Optics Express}, 30(16):29189--29205, 2022.

\bibitem{bianco2022deep}
Vittorio Bianco, Mattia~Delli Priscoli, Daniele Pirone, Gennaro Zanfardino,
  Pasquale Memmolo, Francesco Bardozzo, Lisa Miccio, Gioele Ciaparrone, Pietro
  Ferraro, and Roberto Tagliaferri.
\newblock Deep learning-based, misalignment resilient, real-time fourier
  ptychographic microscopy reconstruction of biological tissue slides.
\newblock {\em IEEE Journal of Selected Topics in Quantum Electronics},
  28(4):1--10, 2022.

\bibitem{holloway2017savi}
Jason Holloway, Yicheng Wu, Manoj~K Sharma, Oliver Cossairt, and Ashok
  Veeraraghavan.
\newblock Savi: Synthetic apertures for long-range, subdiffraction-limited
  visible imaging using fourier ptychography.
\newblock {\em Science advances}, 3(4):e1602564, 2017.

\bibitem{horstmeyer2016diffraction}
Roarke Horstmeyer, Jaebum Chung, Xiaoze Ou, Guoan Zheng, and Changhuei Yang.
\newblock Diffraction tomography with fourier ptychography.
\newblock {\em Optica}, 3(8):827--835, 2016.

\bibitem{zuo2020wide}
Chao Zuo, Jiasong Sun, Jiaji Li, Anand Asundi, and Qian Chen.
\newblock Wide-field high-resolution 3d microscopy with fourier ptychographic
  diffraction tomography.
\newblock {\em Optics and Lasers in Engineering}, 128:106003, 2020.

\bibitem{wu2021imaging}
Daixuan Wu, Jiawei Luo, Guoqiang Huang, Yuanhua Feng, Xiaohua Feng, Runsen
  Zhang, Yuecheng Shen, and Zhaohui Li.
\newblock Imaging biological tissue with high-throughput single-pixel
  compressive holography.
\newblock {\em Nature Communications}, 12(1):1--12, 2021.

\bibitem{lohmann_spacebandwidth_1996}
Adolf~W. Lohmann, Rainer~G. Dorsch, David Mendlovic, Carlos Ferreira, and Zeev
  Zalevsky.
\newblock Space–bandwidth product of optical signals and systems.
\newblock {\em Journal of the Optical Society of America A}, 13(3):470, March
  1996.

\bibitem{ou2013quantitative}
Xiaoze Ou, Roarke Horstmeyer, Changhuei Yang, and Guoan Zheng.
\newblock Quantitative phase imaging via fourier ptychographic microscopy.
\newblock {\em Opt. Lett.}, 38(22):4845--4848, Nov 2013.

\bibitem{zheng2016fourier}
Guoan Zheng.
\newblock {\em Fourier ptychographic imaging: a MATLAB tutorial}.
\newblock Morgan \& Claypool Publishers, 2016.

\bibitem{bian2013adaptive}
Zichao Bian, Siyuan Dong, and Guoan Zheng.
\newblock Adaptive system correction for robust fourier ptychographic imaging.
\newblock {\em Optics express}, 21(26):32400--32410, 2013.

\bibitem{ou2014epfr}
Xiaoze Ou, Guoan Zheng, and Changhuei Yang.
\newblock Embedded pupil function recovery for fourier ptychographic
  microscopy.
\newblock {\em Opt. Express}, 22(5):4960--4972, Mar 2014.

\bibitem{maiden_improved_2009}
Andrew~M. Maiden and John~M. Rodenburg.
\newblock An improved ptychographical phase retrieval algorithm for diffractive
  imaging.
\newblock {\em Ultramicroscopy}, pages 1256--1262, September 2009.

\bibitem{park2021review}
Jongchan Park, David~J Brady, Guoan Zheng, Lei Tian, and Liang Gao.
\newblock Review of bio-optical imaging systems with a high space-bandwidth
  product.
\newblock {\em Advanced Photonics}, 3(4):044001, 2021.

\bibitem{pan2020high}
An~Pan, Chao Zuo, and Baoli Yao.
\newblock High-resolution and large field-of-view fourier ptychographic
  microscopy and its applications in biomedicine.
\newblock {\em Reports on Progress in Physics}, 83(9):096101, 2020.

\bibitem{mudry2012strucyured}
E.~Mudry, K.~Belkebir, J.~Girard, J.~Savatier, E.~Le~Moal, C.~Nicoletti,
  M.~Allain, and A.~Sentenac.
\newblock Structured illumination microscopy using unknown speckle patterns.
\newblock {\em Nature Photonics}, 6(5):312--315, 2012.

\bibitem{dong2014high}
Siyuan Dong, Pariksheet Nanda, Radhika Shiradkar, Kaikai Guo, and Guoan Zheng.
\newblock High-resolution fluorescence imaging via pattern-illuminated fourier
  ptychography.
\newblock {\em Optics express}, 22(17):20856--20870, 2014.

\bibitem{dong2015incoherent}
Siyuan Dong, Pariksheet Nanda, Kaikai Guo, Jun Liao, and Guoan Zheng.
\newblock Incoherent fourier ptychographic photography using structured light.
\newblock {\em Photonics Research}, 3(1):19--23, 2015.

\bibitem{song2019super}
Pengming Song, Shaowei Jiang, He~Zhang, Zichao Bian, Chengfei Guo, Kazunori
  Hoshino, and Guoan Zheng.
\newblock Super-resolution microscopy via ptychographic structured modulation
  of a diffuser.
\newblock {\em optics letters}, 44(15):3645--3648, 2019.

\bibitem{guo2018-13fold}
Kaikai Guo, Zibang Zhang, Shaowei Jiang, Jun Liao, Jingang Zhong, Yonina~C.
  Eldar, and Guoan Zheng.
\newblock 3-fold resolution gain through turbid layer via translated unknown
  speckle illumination.
\newblock {\em Biomedical Optics Express}, 9(1):260--275, 2018.

\bibitem{cheng2019sparse}
Haobo Cheng, Xin Chen, Yongfu Wen, Huaying Wang, and Hui Li.
\newblock Sparse aperture arrangement for periodic led array sampling in a
  fourier ptychography microscopy system.
\newblock {\em Applied Optics}, 58(32):8791--8801, 2019.

\bibitem{mao2020efficient}
Haifeng Mao, Xiaohui Wu, Jufeng Zhao, Guangmang Cui, and Jinxing Hu.
\newblock An efficient fourier ptychographic microscopy method based on
  optimized pattern of led angle illumination.
\newblock {\em Biomedical Optics Express}, 5:102920, 2020.

\bibitem{zhang2015self}
Yongbing Zhang, Weixin Jiang, Lei Tian, Laura Waller, and Qionghai Dai.
\newblock Self-learning based fourier ptychographic microscopy.
\newblock {\em Optics express}, 23(14):18471--18486, 2015.

\bibitem{bian2014content}
Liheng Bian, Jinli Suo, Guohai Situ, Guoan Zheng, Feng Chen, and Qionghai Dai.
\newblock Content adaptive illumination for fourier ptychography.
\newblock {\em Optics letters}, 39(23):6648--6651, 2014.

\bibitem{tian_computational_2015}
Lei Tian, Ziji Liu, Li-Hao Yeh, Michael Chen, Jingshan Zhong, and Laura Waller.
\newblock Computational illumination for high-speed in vitro {Fourier}
  ptychographic microscopy.
\newblock {\em Optica}, 2(10):904--911, October 2015.
\newblock Publisher: Optica Publishing Group.

\bibitem{tian2014multiplexed}
Lei Tian, Xiao Li, Kannan Ramchandran, and Laura Waller.
\newblock Multiplexed coded illumination for fourier ptychography with an led
  array microscope.
\newblock {\em Biomedical optics express}, 5(7):2376--2389, 2014.

\bibitem{dong2014spectral}
Siyuan Dong, Radhika Shiradkar, Pariksheet Nanda, and Guoan Zheng.
\newblock Spectral multiplexing and coherent-state decomposition in fourier
  ptychographic imaging.
\newblock {\em Biomedical optics express}, 5(6):1757--1767, 2014.

\bibitem{yeh2015experimental}
Li-Hao Yeh, Jonathan Dong, Jingshan Zhong, Lei Tian, Michael Chen, Gongguo
  Tang, Mahdi Soltanolkotabi, and Laura Waller.
\newblock Experimental robustness of fourier ptychography phase retrieval
  algorithms.
\newblock {\em Optics express}, 23(26):33214--33240, 2015.

\bibitem{bostan2018accelerated}
Emrah Bostan, Mahdi Soltanolkotabi, David Ren, and Laura Waller.
\newblock Accelerated wirtinger flow for multiplexed fourier ptychographic
  microscopy.
\newblock In {\em 2018 25th IEEE International Conference on Image Processing
  (ICIP)}, pages 3823--3827. IEEE, 2018.

\bibitem{horstmeyer2016standardizing}
Roarke Horstmeyer, Rainer Heintzmann, Gabriel Popescu, Laura Waller, and
  Changhuei Yang.
\newblock Standardizing the resolution claims for coherent microscopy.
\newblock {\em Nature Photonics}, 10(2):68--71, 2016.

\end{thebibliography}

\end{document}